# Semiconductor to metal transition in SWNTs caused by interaction with gold and platinum nanoparticles

Rakesh Voggu[1], Shrinwantu Paul[1], Swapan K. Pati[1,2], C.N.R. Rao[1,*]

[1]Chemistry and Physics of Materials Unit and [2]Theoretical Science Unit,

Jawaharlal Nehru Centre for Advanced Scientific Research,

Jakkur P.O., Bangalore -560 064, India.

E-mail: cnrrao@jncasr.ac.in

Fax: (+91)80-22082766

**Abstract:**

Single-walled carbon nanotubes (SWNTs) have been coated with gold and platinum nanoparticles either by microwave treatment or by the click reaction and the Raman spectra of these SWNT-metal nanoparticle composites have been investigated. Analysis of the G bands in the Raman spectra shows an increase in the proportion of metallic SWNTs on attachment with metal nanoparticles. This conclusion is also supported by the changes observed in the RBM bands. Ab-initio calculations reveal that semiconductor-metal transition occurs in SWNTs due to Columbic charge transfer between the metal nanoparticles and the semiconducting SWNTs.

---

[*] Corresponding author: cnrrao@jncasr.ac.in

**Introduction:**

Single-walled carbon nanotubes (SWNTs) exhibit fascinating electronic, chemical and mechanical properties with a wide range of possible applications [1,2]. As prepared SWNTs are mixtures of semiconducting and metallic nanotubes, the latter being around 33% depending on conditions of the preparation. Many of the applications of SWNTs depend on whether they are semiconducting or metallic. Several attempts are being made to separate semiconducting and metallic nanotubes [3-5]. An equally important aspect is to examine whether the electronic structure of SWNTs is affected by external doping and other means. Thus, it has been found recently that electron or hole doping increases the Raman frequency of the G band, accompanied by decrease in the line width [6]. Electrochemical doping also shifts the Raman G band (1567 $cm^{-1}$) depending on whether nanotubes are semiconducting or metallic [7]. The frequency shifts of SWNTs are clearly associated with the shifts in the Fermi level. Shin et. al.[8] have shown recently that solvents with electron -donating and -withdrawing groups alter the electronic structure of SWNTs. Those with an electron withdrawing group like nitrobenzene, deplete metallic SWNTs while solvents with electron donating group like aniline, donate electrons to both metallic and semiconducting SWNTs.

We have been interested in examining how the electronic structure of SWNTs are affected by attaching metal nanoparticles to them. In order to do this, we have coated SWNTs with nanoparticles of gold and platinum by microwave treatment [9]. We have also covalently attached gold nanoparticles to the SWNTs through the click reaction [10]. We have studied the Raman spectra of SWNTs attached to the metal nanoparticles in comparison with the spectrum of pure SWNTs. We have made use of the G-band in the Raman spectra to obtain the relative proportions of the semiconducting and metallic species. We have also used the radial breathing mode (RBM) bands in the Raman spectra to affirm the results obtained form the analysis of G-bands. The results indeed reveal a considerable increase in the metallic SWNTs on interaction with metal nanoparticles. Ab-initio calculations on Au and Pt fragment coated SWNT composites also show the occurrence of semiconductor to metal transition in SWNTs on interaction with metal nanoparticles.

**Experimental Section:**

SWNTs prepared by arc discharge using the $Y_2O_3$+Ni catalyst were purified by acid and hydrogen treatment following the procedure reported in the literature [11]. The purified SWNTs were treated with a mixture of sulphuric and nitric acids under mild conditions. To coat the SWNTs by the metal nanopraticles, 1mg of nanotubes was taken in a 5ml Teflon-lined autoclave to which 1 mL of chloroauric acid (10mM, $HAuCl_4$) or chloroplatinic acid (10mM, $H_2PtCl_6$) and 1 mL of ethylene glycol were added. The autoclave was placed in a microwave oven equipped with a 700 W magnetron operating at 2.45 GHz for about 30min. The products were subjected to centrifugation, filtration and thorough washing with deionized water to remove loosely bound nanoparticles and ethylene glycol. Gold nanoparticles of 3 and 12nm prepared by literature procedures [12, 13] were covalently linked to SWNTs by the click reaction [10] by the following procedure. SWNTs were first functionalized by the 4-azidobutane-1-amine by reacting acid-treated SWNTs with thionyl chloride, followed by reaction with 4-azidobutane-1-amine. These SWNTs were reacted with Au nanocrystals capped by hex-5-yne-1-thiol in the presence of $CuSO_4$ and sodium ascorbate as a catalyst. The occurrence of the click reaction was verified by infrared spectroscopy from the absence of the azide and acetylenic stretching bands.

**Results and Discussion:**

In figures 1 (a) and (b) we show the TEM images of the SWNTs bundles coated with gold and platinum nanoparticles obtained by microwave treatment. The TEM images show the presence of uniform coatings on the SWNTs. The average diameters of the gold and platinum nanoparticles are around 3 and 2.5 nm respectively. Figures 1(c) and (d) show the TEM images of SWNTs with covalently attached gold nanoparticles of average diameters of 3 and 12 nm through click chemistry.

Figure 2 (a) shows the Raman G band of SWNTs while Figures 2 (b) and (c) shows the G-bands of SWNTs coated with ~3 nm gold nanoparticles by microwave treatment and click reaction respectively. The G band had been decomposed into four or six Lorentzians in the literature [6,14,15]. We first fitted the G bands using four Lorentzians centered at 1510, 1540, 1563 and 1580 cm$^{-1}$, of which first two are due to metallic species, while the 1563cm$^{-1}$ band has contributions from both the metallic and the semiconducting species. The band at 1580 cm$^{-1}$ is mainly due to the semiconducting species. For the purpose of calculating the ratio of the metallic to the semiconducting species, we have used the ratio of areas of the 1540 (metallic, M) and 1580cm$^{-1}$ (semiconducting, S) bands. The M/S ratios so obtained for SWNTs coated with 3 nm gold nanoparticles by microwave treatment and click reaction are 0.61 and 0.72 respectively, compared to ~0.4 in the case of pure SWNTs. These values indicate a 40-80 % increase in the proportion of the metallic SWNT species on interaction with the 3nm gold nanoparticles. With 12 nm Au particles, the increase in the proportion of the metallic species is upto 13%. SWNTs coated with 2.5 nm Pt particles by microwave treatment gave a M/S ratio of 0.73. On fitting the G bands to six Lorentzians centered at 1515, 1540, 1555, 1570, 1580 and 1595 cm$^{-1}$, of which the bands at 1515, 1540 and 1570 cm$^{-1}$ are due to the metallic species, the M/S values showed an increase in the proportion of the metallic species by 20-100% on attachment of 3nm gold or platinum particles by either two methods. We also fitted the 1540cm$^{-1}$ band due to the metallic species to the Fano band shape [8] and obtained the M/S ratios. This analysis showed a 30-40% increase in the metallic species on attaching 3nm Au nanoparticles to the SWNTs. The D-band shows a small increase on attachment of metal particles, but this is not too relevant to the present study.

Pure SWNTs show two RBM bands in the Raman spectra at 145 cm$^{-1}$ and 160 cm$^{-1}$ when excited with 632 nm laser radiation, as can be seen from figure 3(a). Based on Kataura's plot, we assign the band at 145 cm$^{-1}$ due to semiconducting species (S) and the band at 160 cm$^{-1}$ to metallic species (M).[16] Accordingly, excitation by a 514 nm laser source exclusively showed only the band corresponding to the semiconducting species at 145 cm$^{-1}$.[17] Figures 3 (b) and (c) show the RBM bands of SWNTs covered by 3nm Au particles by microwave treatment and click reaction respectively. We see a clear increase in the M/S ratios, obtained by taking the ratios of band areas. The ratios are higher in the case of the 3nm Au particles compared to those with the 12nm gold particles. The smaller Au particles get coated better and are also chemically more active. We found that the Pt particles also cause an increase in the M/S ratio. The results from the RBM bands confirm that there is a definitive increase in the proportion of the metallic species, on interaction of SWNTs with metal nanoparticles. It must however be noted that RBM bands are not suited to obtain quantitative changes in M and S bands unlike the G band.

In Figure 4 we show plots of the percentage increase in the metallic SWNT species found by us on different SWNT samples attached to Au (Pt) nanoparticles and by different methods of analysis of the G bands. This figure clearly establishes that the metal nanoparticles cause an increase in the proportion of the metallic SWNT species. Preliminary studies show that the intensities of the interband transitions show same changes. Thus, the intensity of the $M_{11}$ band in the electronic spectrum increases relative to the intensity of $S_{22}$ band on attachment of SWNTs with gold nanoparticles.

To understand the above experimental results, we have performed ab-initio studies on a variety of SWNT-Au and SWNT-Pt periodic composites, such as, gold and platinum wires, tubes and clusters adsorbed over the metallic as well as semiconducting SWNTs. All the electronic structure calculations were preformed with the SIESTA package [18] at a GGA-PBE level using a DZP basis set for all atoms. A real space mesh cut off of 150 Ry and a Monkhorst [19] k-point grid of 1x1x60 were used with complete atomic relaxations. The calculations were carried out in supercells, chosen such that the interactions between the systems in directions other than the direction in which the system is periodic were negligible. In the case where we considered metal clusters, we further ensured that the interaction between neighboring metal clusters along the direction of periodicity of the system was negligible. All the initial guess systems were fully relaxed to determine the lowest energy structures. The initial guess for the Au and Pt clusters were modeled by Sutton-Chen 12-6 potentials [20].

A single chain of Au and Pt atoms periodically arranged on the surface of a (8,0) SWNT failed to sufficiently alter the electronic structure, although the semiconducting gap of Au-SWNT and Pt-SWNT composites reduced in proportions compared to the pristine (8,0) nanotube. Interestingly, however, the Metal-SWNT system where the entire curved surface area of a semiconducting (8,0) SWNT is covered with a cylindrical metal monolayer is metallic. Furthermore, such extensive interactions between the metal atoms and the SWNT modify the electronic structure of the SWNTs as well, rendering them metallic as shown in the projected DOS plot in figure 5(a). The density of states (DOS) projected out on the SWNT for both the Au-SWNT and Pt-SWNT show that there is sufficient electronic density around the Fermi energy. In the optimized structures, the inter-atomic distances were found to be 2.73 Å for Au and 2.65 Å for Pt, and the difference between the radius of the cylindrical metal monolayers and the radius of the SWNT is 2.84 Å for the Au-SWNT system and 2.82 Å for the Pt-SWNT system. The deviation from the circular tube as measured by the aspect ratio ($r_{max}/r_{min}$) [21] in this case is less than 1%.

To understand the effect of metal clusters adsorbed on to the external curvature of the electronic structure of SWNTs, we considered Au and Pt clusters of various nuclearities adsorbed on the SWNTs. We present here the results of our calculations on Au-SWNT composites involving gold clusters of nuclearity 40, obtained by combining four unit cells of the (8,0) SWNT to ensure that consecutive clusters do not interact appreciably with one another along the direction of periodicity. The $Au_{40}$ cluster was placed such that one vertex of the cluster was at a distance 3.0 Å from the nearest carbon atom. Interestingly, the $Au_{40}$ cluster nearly retains its symmetrical structure, with the nearest proximity of the cluster to the curved surface of the SWNT being 2.4 Å. The optimized electronic structure shows that the DOS projected onto the SWNT had sufficient density around the Fermi energy indicating that the semiconducting SWNT transforms into a metallic one(Figure 5(b)). Even Au clusters of smaller nuclearity are found to transform the semiconducting SWNTs into metallic ones. Upon optimization, Au clusters of various nuclearities relax to highly symmetrical structures similar to the case with the $Au_{40}$ cluster, giving rise to a semiconductor to metal transition. In contrast, the metallic SWNTs do not lead to any new phases for all these cases.

In order to understand the mechanism of semiconductor metal transition, we have calculated the energy and charge densities of the composites by varying the distance between the Au fragment and SWNT. The energy of interaction was found to change inversely with the distance between Au and SWNT, clearly indicating that such transition is due to Coulombic interactions. Interestingly, such interactions are due to resonance between Au plasmon band and the excitation continuum of SWNTs. From the charge density analysis, we also find that the electron density at the outermost valence orbitals of Au reduces while that at the orbitals of carbon in closest proximity to the Au fragments decreases. Thus our results indicate that such a phase transition is due to direct charge transfer between Au and Pt particles and the SWNT.

**Conclusion:**

The present Raman studies established that there is an increase in the proportion of the metallic SWNTs on interaction with Au and Pt nanoparticles. In other words, the metal nanopaticles cause a semiconductor to metal transition in the SWNTs-nanoparticle composites. These results find support from theoretical calculations which reveal that the interaction of SWNTs with small and reasonably large Au and Pt clusters, as well as cylindrical monolayers of these metals, leads to the modification of the electronic structure resulting in semiconductor to metal transition of the SWNTs. The primary interaction responsible for electronic transition is Coulombic charge transfer between the metal particles and SWNT surface. The structures we have considered span a wide range of possibilities of the Au-SWNT composites, all of which show a behaviour with the experimental results. It is noteworthy that the metallic SWNTs do not transform into semiconducting nanotubes upon interaction with these metal particles.

**Acknowledgements:** One of us (R.V.) thanks the University Grants Commission for a fellowship

Figure captions

Figure 1: TEM images of SWNTs coated with (a) gold and (b) platinum nanoparticles by microwave treatment and of SWNTs covalently linked to (c) 3nm and (d) 12nm gold nanoparticles by the click reaction.

Figure 2: G-bands in the Raman spectra of (a) pure SWNTs and of SWNTs coated with 3nm gold nanoparticles (b) by microwave treatment and (c) by the click reaction. S and M stands for semiconducting and metallic species respectively.

Figure 3: RBM bands in the Raman spectra of (a) pure SWNTs and of SWNTs coated with 3 nm gold particles (b) by microwave treatment and (c) by the click reaction.

Figure 4: Percentage increase in the metallic SWNT species on interaction of Au and Pt nanoparticles found with different preparations 1, 2 and 3 and by different methods of analysis of the Raman G band.

Figure 5: (a) DOS plots for cylindrical monolayers adsorbed on the (8,0) SWNT. pDOS for the Au-SWNT composite (thick solid line), Au (dashed line) and pure SWNT (thin solid line). (b) DOS plots for $Au_{40}$ clusters adsorbed on the (8,0) SWNT. pDOS for the Au-SWNT system (thick solid line), Au (dashed line) and SWNT (thin solid line). $E_F$ = -3.26 eV. The vertical line indicate Fermi level in each case.

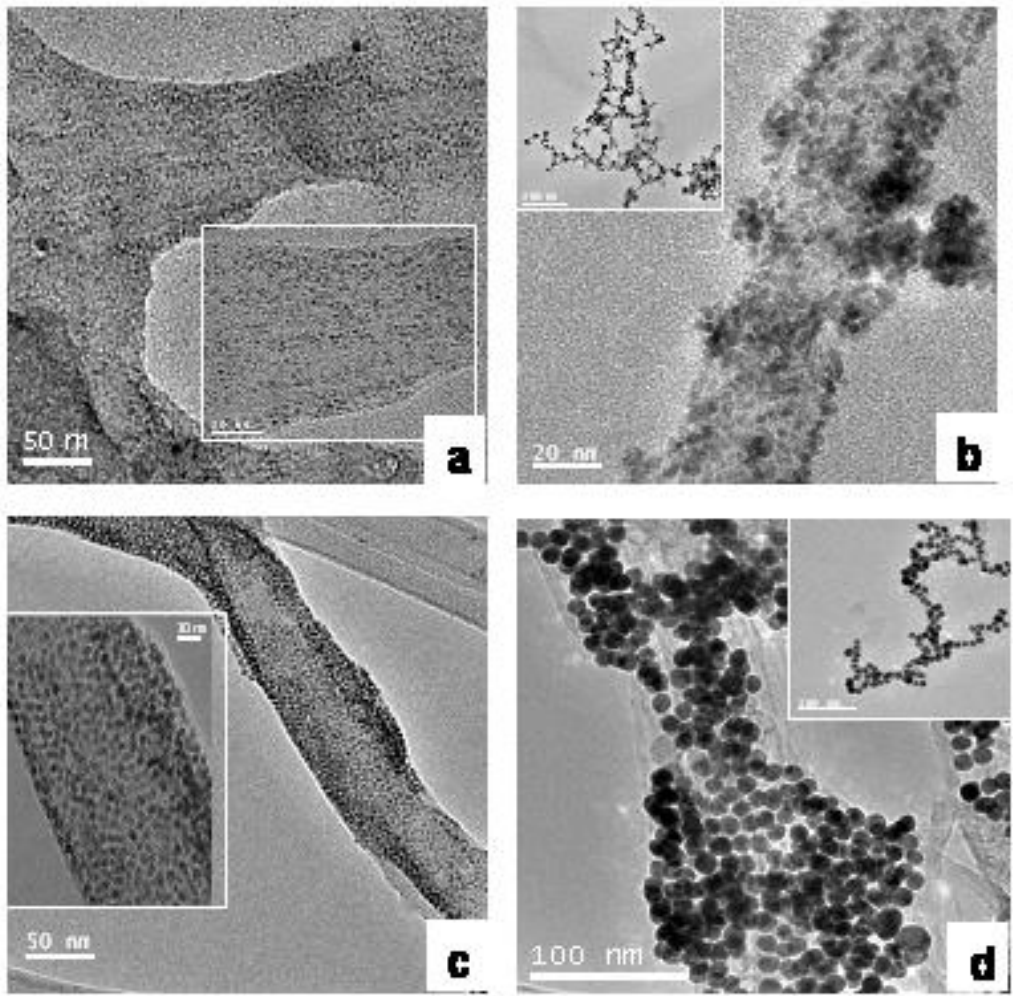

**Figure 1**

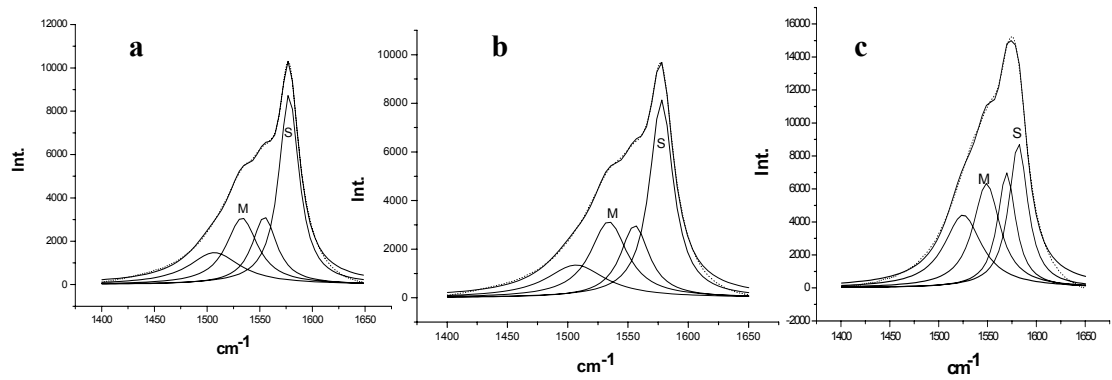

**Figure 2**

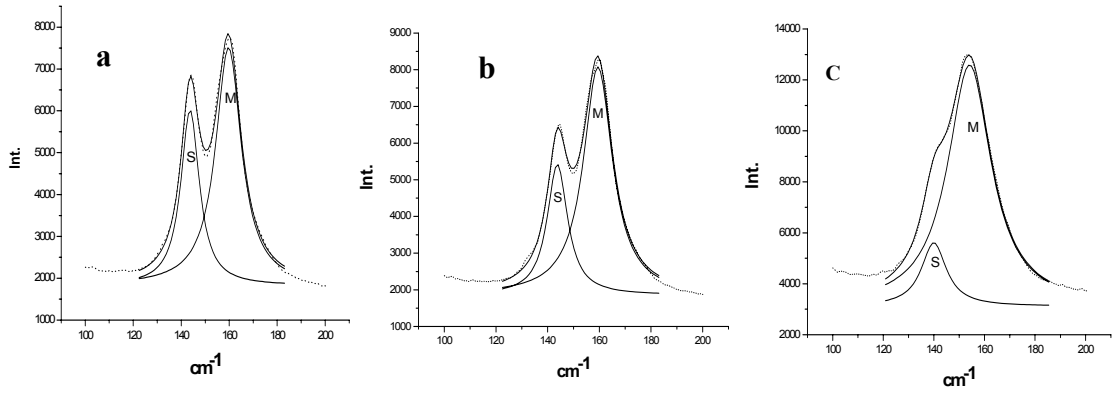

**Figure 3**

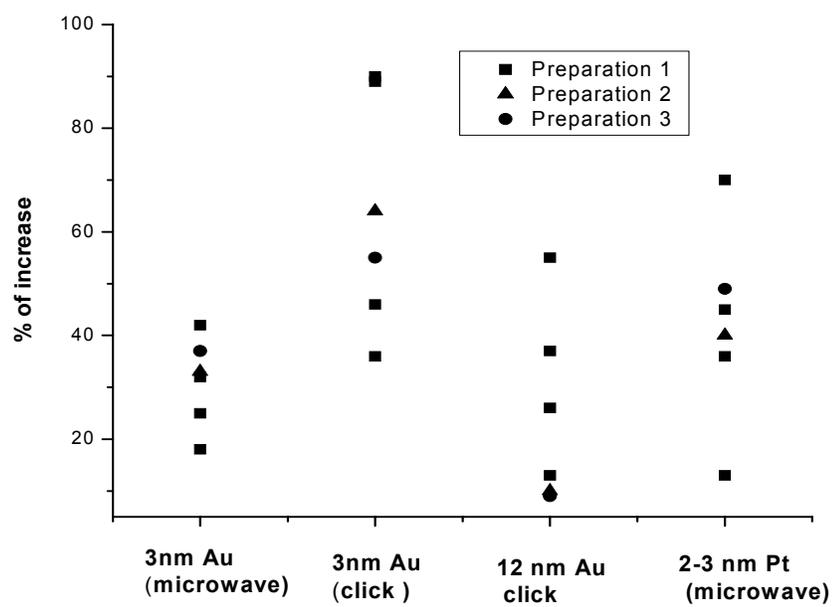

**Figure 4**

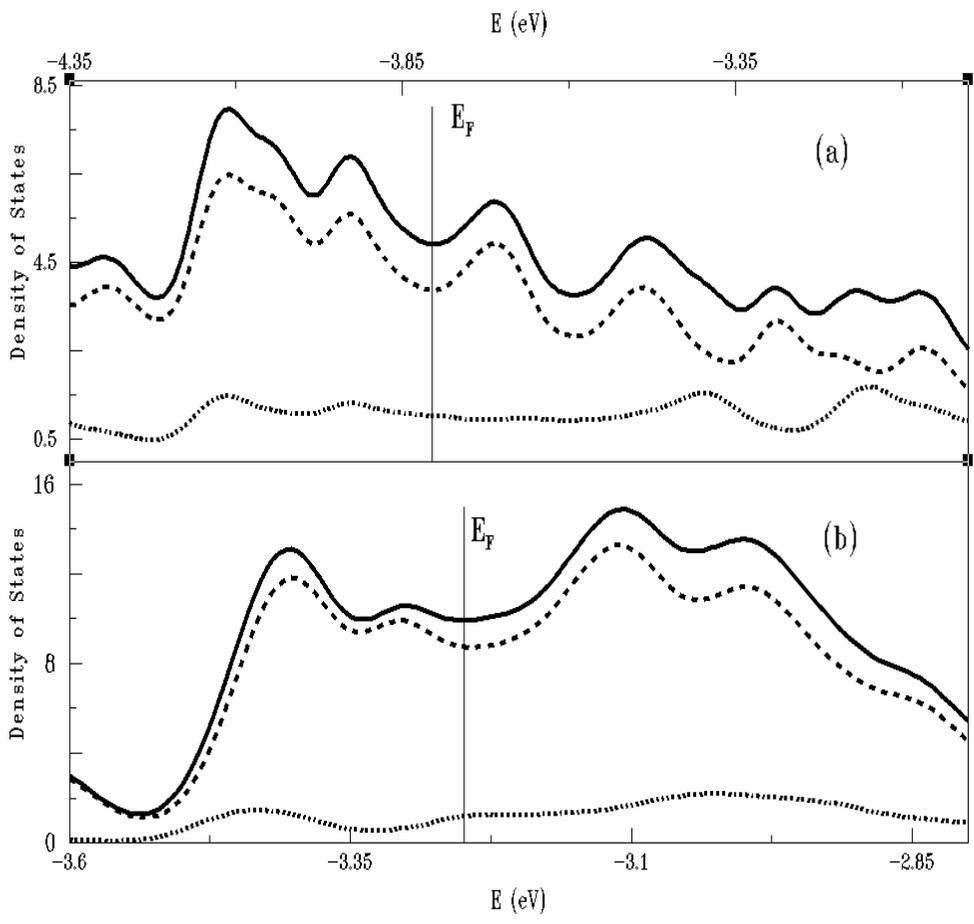

**Figure 5**